\documentclass[10pt]{article}
\usepackage{latexsym}
\usepackage{amssymb}
\usepackage{amsmath}
\usepackage[dvips]{graphicx}

\title{Eigenvalues and eigenvectors of the Laplace operator in $d$ dimensional cut Fock basis}
\author{Piotr Korcyl\thanks{e-mail address: korcyl@th.if.uj.edu.pl} \\ \small{\emph{M. Smoluchowski Institute of Physics, Jagiellonian
University}} \\ \small{\emph{Reymonta 4, 30-059 Krak\'{o}w,
Poland}}}
\date{\today}

\begin{document}

\maketitle

\begin{abstract}
We present exact expressions for the eigenvalues and eigenvectors of the $d$ dimensional Laplace
operator in a cut Fock basis.
\end{abstract}

\section*{Introduction}

The ultimate goal of studies involving Supersymmetric Yang-Mills Quantum Mechanics (SYMQM) is the solution
of the system defined in \mbox{$D=9+1$} space-time dimensions with $SU(N)$ gauge group for any $N$. The latter appears
in such problems as the quantization of a supermembrane \cite{dewit+hoppe+nicolai}\cite{dewit+luscher+nicolai} or the description
of dynamics of $D0$ branes, one of the kinds of objects present in M-theory \cite{banks+fischler+shenker+susskind}\cite{taylor}.
It should not be surprising that finding such solution is a difficult task. Therefore one tries to first
investigate the lower dimensional versions of SYMQM.
Among them, the $D=1+1$ SYMQM is the simplest one \cite{claudson}. It is described by position $\phi^m$
and momentum $\pi^m$ bosonic operators and, $\lambda^m$ and $\bar{\lambda}^m$ fermionic operators.
The index $m$ labels the color degrees of freedom, $m=1,\dots,\mathcal{D}$, where $\mathcal{D}$ is the dimension of the adjoint representation of the gauge group.
Although the $D=2$ SYMQM Hamiltonian is free, its solutions
are not trivial due to the singlet constraint which represents the dimensionally reduced Gauss law. In other words,
the physical Hilbert space of $D=2$ SYMQM is composed of states invariant under the gauge transformations.

A convenient method for analyzing such systems was proposed
few years ago by Wosiek \cite{wosiek} (for a recent introductory article see \cite{korcyl3}). It was designed
mainly as a fully non-perturbative, numerical method for deriving
the spectra of SYMQM.
One of its advantages is that it treats fermions and bosons on an equal footing which made
the calculations in all fermionic sectors possible. Moreover, it can handle any gauge group in any
dimension. The main idea is to calculate the matrix elements of the Hamiltonian operator in a Fock basis composed of
states containing less than $N_B$ quanta. $N_B$ is called the cutoff. Such finite matrix is then diagonalized
numerically in order to obtain the approximated eigenenergies.
A suitable choice of the cutoff, preserving the gauge symmetry as well
as the rotational symmetry, enables one to obtain many numerical results \cite{wosiek1}\cite{wosiek2}\cite{wosiek4}\cite{wosiek3}.
The cut Fock space approach turned out also to be a suitable tool for analytic investigations.

In this short paper we analyze the $D=2$ SYMQM models with the $SO(d)$ gauge group.
In the language of SYMQM they can be
interpreted as an extremely simplified $d$-dimensional model with only one color or, equivalently,
as a model with $d$ colors transforming in the adjoint representation of the $SO(d)$ group.
Due to the simplicity of the $SO(d)$ symmetry group we are able to derive analytic expressions for the cut spectrum and eigenstates.
Consequently, we check that the infinite cutoff limit of the cut solutions
is given by the well-know Bessel functions \cite{abramowitz}\cite{bateman},
which are solutions of the free
$d$-dimensional Schr\"odinger equation carrying zero angular momentum.
We expect that such analytic solutions can be used as a starting point for
the study of systems with several colors transforming with the $SU(N)$ group.
Moreover, once known for any $d$ or $N$ they could be instructive
in the further study of physically interesting 'large-d' or 'large-N' limits.

We also discuss the physical interpretation of the cutoff imposed on the Fock basis. We show that it corresponds to a specific
discretization of position and momentum spaces. Afterwards, we briefly analyze the scaling law, i.e. the dependence
of the eigenenergies on the labeling index. Such law was first investigated
by Wosiek and Trze\-trze\-le\-wski \cite{maciek2}\cite{maciek1} in one spatial dimension and
was used as a handle to correctly
differentiate localized states from the
non-localized ones within our numerical approach.
In this note we discuss its validity in higher dimensional spaces.

We summarize our results in the last section.


\section*{The system}

Traditionally, in the $D=2$ SYMQM models $\phi^m$ and $\pi^m$,
as well as, $\lambda^m$ and $\bar{\lambda^m}$
transform in the adjoint representation of the $SU(N)$ gauge symmetry.
In this paper we concentrate
on the $SO(d)$ gauge symmetry, and
therefore, we assume $\phi^m$ and $\lambda^m$ to be in the adjoint representation of $SO(d)$.
The SYMQM Hamiltonian is supplemented by the singlet constraint, hence we search for solutions carrying zero angular momentum.
The Hamiltonian reads
\begin{equation}
H = \frac{1}{2} \sum_{m=1}^{d} \pi^m \pi^m.
\end{equation}
One introduces standard creation and annihilation operators,
\begin{equation}
a^{\dagger m} = \frac{1}{\sqrt{2}}(\phi^m - i \pi^m), \qquad
a^m = \frac{1}{\sqrt{2}}(\phi^m + i \pi^m)
\end{equation}
fulfilling the commutation relation,
\begin{equation}
[a^n, a^{\dagger m}] = \delta^{m n},
\end{equation}
and rewrite $H$ as
\begin{equation}
H = a^{\dagger m} a^m + \frac{d}{2} - \frac{1}{2} a^m a^m - \frac{1}{2} a^{\dagger m}a^{\dagger m},
\end{equation}
where the summation from 1 to $d$ over repeated indices is understood from now on.

The $SO(d)$ group possess only one invariant symmetric tensor, namely the Kronecker delta.
Hence, the algebra of invariant under rotations creation operators reads
\begin{align}
\begin{split}
[a^m a^m, a^{\dagger n} a^{\dagger n}]  &= 2 d + 4 a^{\dagger n} a^n,
\\
[a^{\dagger m} a^m, a^{\dagger n} a^{\dagger n} ] &= 2 a^{\dagger n} a^{\dagger n},
\\
[a^p, a^{\dagger m} a^{\dagger m} ]  &= 2 a^{\dagger p}.
\label{eq. commutation relations}
\end{split}
\end{align}

The basis vectors in the singlet sector can be labeled by a single index $n$, and take the form
\begin{equation}
|n\rangle  = \frac{1}{\mathcal{N}^0_n} \big( a^{\dagger m} a^{\dagger m} \big)^n |0\rangle.
\label{eq. basis states}
\end{equation}
Notice that $n$ is related to the half of the total number of quanta of each type contained in $|n\rangle$, $N_B = 2n$.
For a given cutoff $N_B$ only states $|n \le \lfloor \frac{N_B}{2} \rfloor \rangle$ are present.
There are $\lfloor \frac{N_B}{2} \rfloor+1$ such states, a new one appearing for $N_B$ even.

We calculate the normalization factor in eq.\eqref{eq. basis states} 
by commuting consecutively all
operators $a^q a^q$ through $(a^{\dagger p}a^{\dagger p})^i$, using eqs.\eqref{eq. commutation relations}, so that they annihilate the Fock vacuum.
This gives a recursive relation for $\mathcal{N}^0_n$,
\begin{equation}
\big(\mathcal{N}_n^0\big)^2 = 2n (2n +d -2) \big( \mathcal{N}^0_{n-1} \big)^2, \qquad \big(\mathcal{N}_0^0\big)^2 = 1,
\end{equation}
which is solved by
\begin{equation}
\big(\mathcal{N}_n^0\big)^2 = 4^n  n! \frac{ \Gamma(\frac{d}{2}+n)}{\Gamma(\frac{d}{2})}.
\label{eq. normalization}
\end{equation}

Although we are mainly interested in the scalar sector, we can obtain results for higher representations
without additional effort. As an example, we present
some calculations in the vector sector. The basis states in this sector can be obtained out of
the states given by eq.(\ref{eq. basis states}) by the action of
an additional single creation operator,
\begin{equation}
|n, i \rangle = \frac{1}{\mathcal{N}^1_n} a^{\dagger i} ( a^{\dagger m} a^{\dagger m} )^n |0 \rangle, \qquad i=1, \dots , d,
\label{eq. basis states j=1}
\end{equation}
where
\begin{equation}
(\mathcal{N}_n^1)^2 = 4^n n! \frac{\Gamma(\frac{d}{2}+n+1)}{\Gamma(\frac{d}{2}+1)}.
\label{eq. normalization j=1}
\end{equation}
States given by eq.(\ref{eq. basis states j=1}) are orthogonal in $n$
and $i$: $\langle m,j | n,i \rangle = \delta_{m n} \delta_{j i} \big( \mathcal{N}^1_n \big)^2$.



It is now straightforward to evaluate all elements of the Hamiltonian matrix in the singlet sector,
\begin{align}
\begin{split}
\langle n| H |n\rangle = 2 n + \frac{d}{2}, \quad
\langle n+1| H |n\rangle = \langle n| H |n+1\rangle = -\sqrt{(n+1)(\frac{d}{2}+n)},
\label{eq. hamiltonian matrix elements j=0}
\end{split}
\end{align}
and in the vector sector,
\begin{align}
\begin{split}
\langle n, j| H | n, i \rangle &= \delta_{j i} \big( \frac{d}{2} + 2n+1 \big),
\\
\langle n, j| H | n+1, i \rangle &= \langle n+1, j| H | n, i \rangle = - \delta_{j i} \sqrt{(n+1)(\frac{d}{2}+n+1)}.
\label{eq. hamiltonian matrix elements j=1}
\end{split}
\end{align}

\begin{figure}[!h]
\begin{center}
\input{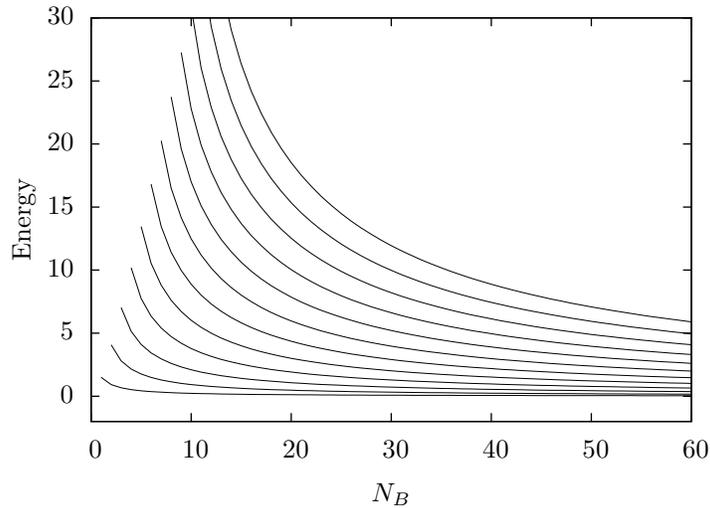}
\caption{Spectrum of the Laplace operator in $d=3$ dimensions in the singlet sector. \label{fig. spectra}}
\end{center}
\end{figure}

This done, we can calculate numerically the spectra. In order to obtain the physical results corresponding
to the infinite cutoff limit we investigate the behavior of the eigenenergies with increasing cutoff for some
finite values of $N_B$. Figure \ref{fig. spectra} contains such results for the 12 lowest eigenenergies in the singlet sector.
Each curve falls slowly to zero as expected from the nonlocalized character of the eigenstates of
the free Hamiltonian \cite{maciek2}\cite{maciek1}\cite{praca_magisterska}. The dependence on the cutoff
will be also discussed in the last section of this note.

\section*{Character method}

There exists an independent way to calculate the number of basis states
transforming according to a given representation of the $SO(d)$ group, which
exploits the orthogonality of characters of irreducible
representations\footnote{The method was suggested by R. Janik}.
We start with a formula for the number of singlets appearing in a product of $N_B$ adjoint representations \cite{doktorat_macka}
\begin{equation}
d^d_{N_B} = \int d_{\mu_{SO(d)}}(\alpha_i) \ 1 \cdot \chi_{N_B}^d(\alpha_i),
\label{eq. charaktery 1}
\end{equation}
where $d_{\mu_{SO(d)}}(\alpha_i)$ is the invariant Haar measure for the $SO(d)$ group parameterized by the set of
angles $\big\{ \alpha_i \big\}_{i=1,\dots,\lfloor d/2 \rfloor}$ running from $0$ to $2 \pi$ and $\chi_{N_B}^d(\alpha_i)$
represents the symmetrized product
of $N_B$ characters of the adjoint representation of the $SO(d)$ group and is equal to \cite{hamermesh}
\begin{equation}
\chi_{N_B}^{d}(\alpha_i) = \sum_{\sum_k k i_k = N_B} \prod_{k=1}^{N_B} \frac{1}{i_k!} \frac{1}{k^{i_k}}
\Big( \chi^d(k \alpha_i) \Big)^{i_k}.
\end{equation}
$1$ in eq.(\ref{eq. charaktery 1}) stands for the trivial character of the invariant representation of $SO(d)$. It can be replaced
by the character of any other representation $R$, in which case, $d^d_{n_B}$ will be equal to the number of times $R$ appears in the
product of $N_B$ adjoint representations.
For $d$ even, $d = 2M$ with $M$ integer, the Haar measure for $SO(d)$ is given by \cite{teoria_grup}
\begin{equation}
d_{\mu_{SO(d)}} = \frac{2^{(M-1)^2}}{\pi^M M!} \prod_{1 \le j < k \le M} \Big( \cos(\alpha_k) - \cos(\alpha_j)\Big)^2,
\end{equation}
and for $d$ odd, $d = 2M+1$,
\begin{equation}
d_{\mu_{SO(d)}} = \frac{2^{M^2}}{\pi^M M!} \prod_{1 \le j < k \le M} \Big( \cos(\alpha_k) - \cos(\alpha_j)\Big)^2 \prod_{m=1}^M \sin^2(\frac{\alpha_m}{2}).
\end{equation}
The characters of the adjoint representation of $SO(d)$ are given by, for $d=2M$ \cite{teoria_grup},
\begin{equation}
\chi^d(\alpha_i) = 2 \sum_{m=1}^M \cos(\alpha_m),
\end{equation}
and for $d=2M+1$,
\begin{equation}
\chi^d(\alpha_i) = 1 + 2 \sum_{m=1}^M \cos(\alpha_m).
\end{equation}

Explicit evaluation of the integrals confirms that eqs. \eqref{eq. basis states}
and \eqref{eq. basis states j=1} form indeed a complete basis of the Hilbert space
in the corresponding sectors.


\section*{Analytic results}

In this section we derive the set of eigenvalues and eigenvectors of the Hamiltonian operator
in the singlet and vector sectors for any finite cutoff $N_{B}$.

At cutoff equal to $N_B$, the basis size in the singlet sector is $n_{max} = \lfloor\frac{N_B}{2}\rfloor +1$.
Let us denote by $I_{n_{max}}(\lambda)$ the characteristic polynomial of the Hamiltonian matrix at this cutoff.
Then, $I_1 = \langle 0|H - \lambda|0\rangle = \frac{d}{2} - \lambda$. Due to the tridiagonal
nature of the Hamiltonian matrix we can write a recursive relation for $I_{n}(\lambda)$,
\begin{align}
I_{n}(\lambda) = \Big( 2 n-2 + \frac{d}{2} - \lambda \Big) I_{n-1}(\lambda) - \Big( (n-1) (n-2 + \frac{d}{2})\Big) I_{n-2}(\lambda).
\end{align}
By changing the variables as $I_{n}(\lambda) \rightarrow n! I_{n}(\lambda)$ we get
a recursive equation which is solved by generalized Laguerre polynomials, i.e.
\begin{equation}
I_{n}(\lambda) = L_{n}^{\frac{d}{2}-1}(\lambda).
\label{eq. wyznacznik}
\end{equation}
These well-known polynomials are defined
as the solutions of the differential equation
\begin{equation}
x y'' + (\alpha +1 -x) y' + n y = 0, \nonumber
\end{equation}
and the orthogonality relation
\begin{equation}
\int_0^{\infty} L_m^{\alpha}(x) L_n^{\alpha}(x) x^{\alpha} e^{-x} dx = \delta_{m n}. \nonumber
\end{equation}
Therefore the eigenvalues of the Hamiltonian operator in the singlet sector in $d$ dimensions at cutoff $N_B$ are given
by the zeros of the Laguerre polynomial $L_{\lfloor \frac{N_B}{2} \rfloor+1}^{\frac{d}{2}-1}(\lambda)$. 
By taking $d=1$ we recover the results of \cite{maciek1}.

An alternative method to derive this result is to consider a general state $|E\rangle$ from the
singlet sector and expand it as
\begin{equation}
|E\rangle = \sum_{n=0}^{\lfloor \frac{N_{B}}{2} \rfloor} a_{n}(E) |n\rangle.
\end{equation}
The eigenequation of $H$ translates into a recursive relation for the $a_n(E)$
coefficients. After extracting an irrelevant constant factor, $a_{n}(E) \rightarrow 2^{-n} a_{n}(E)$, the recursion reads
\begin{equation}
a_{n-1}(E) - \big( 2n + \frac{d}{2} - E \big) a_{n}(E) + (n+1)\big( n + \frac{d}{2} \big) a_{n+1}(E) = 0.
\label{eq. relacje rekurencyjna}
\end{equation}
Eq.\eqref{eq. relacje rekurencyjna} is solved by
\begin{equation}
a_{n}(E) = a_0 \Gamma(\frac{d}{2}) \frac{L^{\frac{d}{2}-1}_n(E)}{\Gamma(n+\frac{d}{2})}, \qquad n=1,\dots,\lfloor \frac{N_B}{2} \rfloor,
\label{eq. eigenvectors j=0}
\end{equation}
%
%
together with the quantization condition $L^{\frac{d}{2}-1}_{\lfloor \frac{N_B}{2} \rfloor+1}(E)=0$, which is equivalent to eq.\eqref{eq. wyznacznik}.

Similarly, in the vector sector the recursive relation for the determinant
of the Hamiltonian matrix
can be solved by
\begin{equation}
I_{n}(\lambda) = L_{n+1}^{\frac{d}{2}}(\lambda).
\end{equation}
Thus, the spectrum in the vector sector for the cutoff $N_B$ is given by the zeros of
the generalized Laguerre polynomials with the index shifted by $1$ comparing to the singlet case,
\begin{equation}
H |E_n^i\rangle = E_n |E^i_n\rangle, \ \textrm{with } E_n \textrm{ such that } L_{\lceil \frac{N_B}{2} \rceil +1}^{\frac{d}{2}}(E_i) = 0, \quad n=1,\dots, \lceil \frac{N_B}{2} \rceil + 1 \nonumber
\end{equation}
where $|E^i_n \rangle$ are the eigenstates of $H$ with the vector index $i$. As
a straightforward consequence of the vector nature of the states $|E^i_n \rangle$
each eigenvalue is $d$ times degenerate.
The corresponding eigenvectors are given as an expansion in the basis
\begin{equation}
|E^i\rangle = \sum_{n=0}^{\lceil \frac{N_{B}}{2} \rceil} a_{n,i}(E) |n,i\rangle.
\end{equation}
with
\begin{equation}
a_{n,i}(E) = a_0 \Gamma(\frac{d}{2}) \frac{L^{\frac{d}{2}}_n(E)}{\Gamma(n+\frac{d}{2}+1)}, \qquad n=1,\dots,\lceil \frac{N_B}{2} \rceil.
\end{equation}

Hence, we obtained simple formulae giving the spectrum of the cut Hamiltonian operator as well as its eigenstates. The dependence on the
cutoff is explicit and it is straightforward to perform the infinite cutoff limit.

One can easily generalize the above expressions for the eigenvalues and eigenvectors transforming under any irreducible representation of the $SO(d)$ group.

\section*{Reconstruction of radial waves}


In this section we confirm the results obtained above by comparing their
infinite cutoff limit with the well-known continuous results.
Before we present the detailed calculations, we make two observations.

The rescaled Hamiltonian of the quantum harmonic oscillator is given by
\begin{equation}
H_{osc} = p^2 + x^2.
\end{equation}
It is evident that $H_{osc}$ is invariant under the transformation
$x \leftrightarrow p$. Therefore, the functional form of
the wave functions in the momentum and position representations is the same up to a constant multiplicative factor.
In fact, both are related by the $d$ dimensional Fourier transform.

Next, the amplitude $a_{n}(E)$ can be interpreted as the $n$-th
wave function of the $d$-dimensional harmonic oscillator in the momentum representation.
To see this, let as denote the plane wave with a wave-vector $\kappa$, $\kappa = \sqrt{E}$, which is the
eigenstate of the free Hamiltonian $H$, by $|\kappa \rangle$, and the
$n$-th state of the $d$-dimensional harmonic oscillator carrying zero angular momentum by $|n\rangle$.
Then, $a_{n}(E)$ is simply the scalar product of the free wave $|\kappa \rangle$ and the harmonic oscillator eigenfunction $|n\rangle$ ,
\begin{equation}
a_n(E) = a_{n}(\kappa^2) = \langle \kappa | n \rangle.
\end{equation}

Thus,
\begin{align}
\psi_n(\kappa) \equiv \langle \kappa|n\rangle &= f(\kappa^2) L_n^{\frac{d}{2}-1}(\kappa^2), \\
\phi_n(r) \equiv \langle r|n\rangle &= g(r^2) L_n^{\frac{d}{2}-1}(r^2),
\end{align}
where $f(\kappa^2)$ and $g(r^2)$ are functions which do not depend on $n$ and could not be determined from the recursion relations.
We can check the above conclusions by explicitly solving the
radial $d$ dimensional harmonic oscillator eigenequation in the position representation,
\begin{equation}
\frac{1}{2} \frac{d^2 \phi_n(r)}{dr^2} + \frac{d-1}{2 r} \frac{d \phi_n(r)}{dr} + (2n + \frac{d}{2}) \phi_n(r) - \frac{1}{2} r^2 \phi_n(r) = 0.
\end{equation}
The solutions to this equation are
\begin{equation}
\phi_n(r) = c_1 e^{-\frac{r^2}{2}} r^{2-d} {}_1 F_1 \big(1 -\frac{d}{2}-\frac{n}{2}, 2 - \frac{d}{2}, r^2 \big) + c_2 e^{-\frac{r^2}{2}} {}_1 F_1 \big(-\frac{n}{2}, \frac{d}{2}, r^2 \big).
\end{equation}
The first term of the solution is singular at $r=0$ for $d>2$, so we set $c_1=0$. By rewriting $\phi_n(r)$ in a more familiar form we get
\begin{equation}
\phi_n(r) = c_2 e^{-\frac{r^2}{2}} L_{n}^{ \frac{d}{2}-1} ( r^2 ),
\end{equation}
and similarly
\begin{equation}
\psi_n(\kappa) = c_2 e^{-\frac{\kappa^2}{2}} L_{n}^{ \frac{d}{2}-1} ( \kappa^2 ),
\end{equation}
where the constant $c_2$ can be fixed by the normalization condition,
\begin{equation}
c_2 = \sqrt{\frac{ n! \Gamma(\frac{d}{2})}{\pi^{\frac{d}{2}} \Gamma(n+\frac{d}{2})}}.
\end{equation}
At this point we can calculate the wave functions in the position
representation of the infinite cutoff limit of solutions eq.(\ref{eq. eigenvectors j=0}). We write
\begin{multline}
\langle r|\kappa\rangle = \sum_{n=0}^{\infty} \langle r|n\rangle \langle n|\kappa\rangle =
\sum_{n=0}^{\infty} \phi_n(r) \psi_n(\kappa) \\
= \mathcal{N} e^{-\frac{\kappa^2+r^2}{2}} \sum_{n=0}^{\infty} \frac{(-1)^n n!}{\Gamma(n+\frac{d}{2})} L_{n}^{ \frac{d}{2}-1} ( r^2 ) L_{n}^{ \frac{d}{2}-1} ( \kappa^2 ),
\end{multline}
with all unimportant constant factors gathered in $\mathcal{N}$ and the $(-1)^n$ factor comeing from the definition of Fourier transform.
Using the known relation for associated Laguerre polynomials \cite{bateman}
we eventually get
\begin{align}
\langle r|\kappa\rangle &= \frac{1}{2} \mathcal{N} \frac{1}{(- \kappa r)^{\frac{d}{2}-1}} J_{\frac{d}{2}-1} \big( - \kappa r \big).
\label{eq. spherical wave solution}
\end{align}
This result indeed coincides with the solutions of the radial part of the Laplace equation,
\begin{equation}
-\Delta \phi(r) - E \phi(r) = \frac{d^2 \phi(r)}{dr^2} + \frac{d-1}{r} \frac{d \phi(r)}{dr} + E \phi(r) = 0.
\end{equation}
given by the well-known Bessel functions,
\begin{equation}
\phi(r) = c_1 r^{1-\frac{d}{2}} J_{\frac{d}{2}-1} (-\sqrt{E} r) + c_2 r^{1-\frac{d}{2}} Y_{\frac{d}{2}-1} (-\sqrt{E} r).
\end{equation}
Requiring an analytic behavior of the solution at $r=0$ fixes $c_2=0$, and after setting $c_1 = \frac{1}{2} \mathcal{N} \kappa^{1-\frac{d}{2}}$ and $E=\kappa^2$,
we recover the solutions of eq.(\ref{eq. spherical wave solution}).
Therefore, our solutions obtained for a finite $N_B$ become, in the infinite cutoff limit, the known Bessel functions.

%
%

\section*{Physical interpretation of the cutoff}

Physically, the introduction of a cutoff limiting the maximal number of quanta contained in the basis states is equivalent
to a specific discretization of the position as well as momentum variables.
This can be demonstrated by considering the eigenvalues
of the cut matrix representations of those operators. They correspond to possible outcomes of a measurement process. Obviously,
in the cut basis the matrices of the $\hat{x}$ and $\hat{p}$ operators are finite and
have as many eigenvalues as there are states in the cut basis. Hence, position as well as momentum spaces are discrete.

In order to be more precise, let us consider gauge invariant position and momentum operators of the form
$X^2=x^m x^m$ and $P^2=p^m p^m$. We have shown in the preceding sections that their spectra consist of the zeros of
Laguerre polynomials. We denote their eigenvalues by $X^2_i$ and $P^2_i$, respectively,
\begin{equation}
X^2_i: \ L_{n}^{\frac{d}{2}-1}(X^2_i)= 0, \qquad P^2_i: \ L_{n}^{\frac{d}{2}-1}(P^2_i) = 0, \qquad i=1,\dots,n. \nonumber
\end{equation}
%
%
%
%
%
%
A well-known upper bound for the largest zero of $L_n^{\alpha}(x)$ is, provided $|\alpha| \ge \frac{1}{4}$, $\alpha >-1$,
\begin{equation}
x_n < \Big( \sqrt{4n + 2\alpha + 2} - \gamma (4n + 2\alpha + 2)^{-\frac{1}{6}} \Big)^2, \nonumber
\end{equation}
where $\gamma = 6^{-\frac{1}{3}} i_1$ and $i_1$ is the smallest zero of the Airy function \cite{szego}.
Hence, for a cutoff $N_B$ the eigenvalues $X_i$ and $P_i$
lie in the interval $\mathcal{I}$,
\begin{equation}
\mathcal{I} = (-\sqrt{4 \big\lceil \frac{N_B}{2} \big\rceil + 3}, \ \sqrt{4 \big\lceil \frac{N_B}{2} \big\rceil + 3} ). \nonumber
\end{equation}
The interval $\mathcal{I}$ of length $\sqrt{N_B}$ contains roughly $\frac{1}{2} N_B$ such eigenvalues. Therefore, one concludes
that $X_i$ and $P_i$ tend to cover the real axis in the limit of infinite cutoff.
We call this limit also a continuum limit.

\section*{Continuum limit}


Up to now, our results suggest that the continuum limit of the $n$-th eigenenergy vanishes,
\begin{equation}
\lim_{N_B\rightarrow\infty} E_n(N_B) = 0,
\label{eq. bad scaling}
\end{equation}
However this is in contradiction with the general expectation
that the spectrum of the kinetic energy operator is continuous, i.e.
\begin{equation}
\lim_{N_B\rightarrow\infty} E_n(N_B) = \frac{1}{2} p^2, \quad p \in \mathbb{R},
\label{eq. scaling requirement}
\end{equation}
In order to avoid the situation of eq.\eqref{eq. bad scaling} one must assume
that the limit in eq.\eqref{eq. scaling requirement} is taken with,
\begin{equation}
n = n(N_B),
\label{eq. scaling}
\end{equation}
which we call \emph{scaling law}, referring to a similar procedure in lattice field theory.
This phenomenon was investigated in the case of one-dimensional momentum operators by Trzetrzelewski \cite{maciek1}.
He was able to derive the scaling properties of the eigenvalues of the momentum
operator as well as of the kinetic energy operator. We report the latter in table \ref{tabela}.
The four cases shown correspond to different parities of the corresponding eigenstates. Such complication
is not present for $d>1$ since the basis states are always even, i.e. build of powers of $a^{\dagger 2}$.
\begin{table}[!h]
\begin{center}
\begin{tabular}{c|cc}
 & $N_B \textrm{ odd}$ & $N_B \textrm{ even}$ \\
\hline
&& \\
$n \textrm{ odd}$ & $E_n(N_B) \approx \frac{\pi^2}{2} \frac{(n-\frac{1}{2})^2}{2N_B+3}$ & $E_n(N_B) \approx \frac{\pi^2}{2} \frac{(n-\frac{1}{2})^2}{2N_B+5}$ \\
&& \\
$n \textrm{ even}$ & $E_n(N_B) \approx \frac{\pi^2}{2} \frac{n^2}{2N_B+5}$ & $E_n(N_B) \approx \frac{\pi^2}{2} \frac{n^2}{2N_B+3}$ \\
&& \\
\end{tabular}
\end{center}
\caption{Scaling laws for the eigenenergies for even and odd $N_B$ and $n$. \label{tabela}}
\end{table}

%

\begin{figure}
\begin{center}
\input{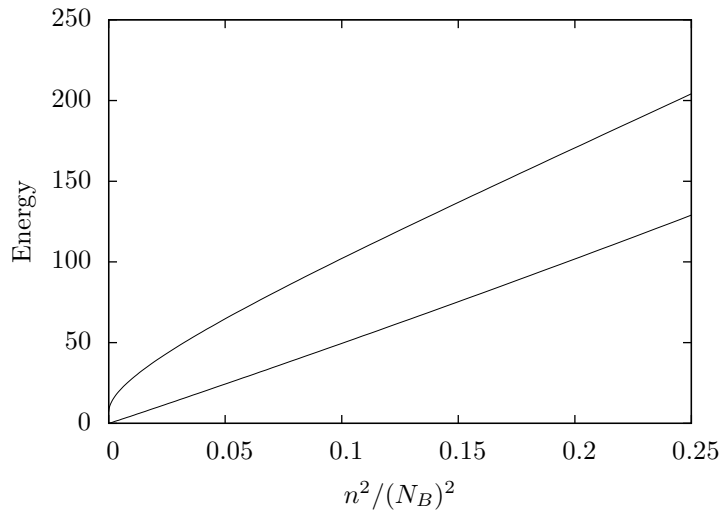}
\caption{Scaling laws for $d=1$ and $d=150$. \label{fig. scaling}}
\end{center}
\end{figure}

It is now straightforward to check the validity of the scaling law in $d>1$ dimensions.
The eigenenergies of the kinetic energy operator being the zeros of appropriate Laguerre polynomials
can be approximated
at large $N_B$ by the formula \cite{abramowitz}\cite{bateman}
\begin{equation}
E_n(N_B) = \big(L^{\frac{d}{2}-1}_{\lfloor \frac{1}{2}N_B \rfloor +1}(E)\big)_n \approx \frac{j_{\frac{d}{2}-1,n}^2}{4\lfloor \frac{1}{2}N_B \rfloor+d} + \dots,
\end{equation}
Hence, we see that the simple scaling $E_n \sim n^2$ is correct as long as the following approximation is true
\begin{equation}
j_{\frac{d}{2}-1,n} \approx \gamma_1(d) n + \gamma_2(d).
\label{eq. approx}
\end{equation}
Obviously, \eqref{eq. approx} will not hold for high $d$. Indeed,
Figure \ref{fig. scaling} shows the dependence of the eigenenergies $E_n(N_B)$ on the labeling number $n$ for a given cutoff \mbox{$N_B=200$}.
Distinct curves correspond to different $d$, namely $d=1$ for the lower one and $d=150$ for the upper one.
Definitely, in the latter case the dependence is no longer linear.
%
%
%
%
%
%
Some particular values of the coefficients $\gamma_1$ and $\gamma_2$ are given in table \ref{tabela 2}.
\begin{table}[!h]
\begin{center}
\begin{tabular}{|c||c|c||c|c|}
\hline
d & $\gamma_1(d)$ & Std. error & $\gamma_2(d)$ & Std. error\\
\hline
1 & $3.14159$ & $< 10^{-10}$ & -1.5708 & $< 10^{-10}$\\
3 & $3.14159$ & $< 10^{-10}$ & 0.0 & $<10^{-10}$ \\
9 & $3.142$ & $4.4 \times 10^{-5}$ & $4.6192$ & $0.0077$ \\
\hline
\end{tabular}
\end{center}
\caption{Fitted values of parameters $\gamma_1(d)$ and $\gamma_2(d)$ for $d=1,3,9$. \label{tabela 2}}
\end{table}

\section*{Conclusions}

In this note we used the cut Fock space approach to investigate the system
which can be interpreted as a free quantum particle in $d$ dimensions.
Although solutions to such problem are well known in the continuum ($N_B=\infty$),
they were unknown for finite $N_B$.

Our approach was guided by the numerical algorithm proposed for solving supersymmetric
Yang-Mills quantum mechanics in the Hamiltonian formulation.
The study of this particular system was motivated by a similar
problem with the $SO(d)$ gauge group replaced by the $SU(N)$ group \cite{korcyl4}\cite{korcyl1}.

We started by describing
the construction of basis states with
the use of gauge invariant creation operators. Then, we explicitly derived the Hamiltonian matrix. Due to its simple form
it was possible to obtain exact expressions for the eigenvalues and eigenvectors. The formulae are parameterized by the cutoff and
hence, are valid for any finite cutoff. Namely, the spectrum at finite cutoff is given by the
zeros of an appropriate Laguerre polynomial, whereas in
the continuum it corresponds to the positive real axis.

In order to confirm the correctness of our solutions,
we calculated their wave functions in the position representation in the continuum limit. Indeed, it was possible to show
that such wave functions are equivalent to the well known
solutions given in terms of Bessel functions.

Next, we discussed the physical interpretation of the cutoff. We argued that a finite cutoff corresponds to a discretization
of both position and momentum space. In the infinite cutoff limit continuum results are recovered.

Finally, we observed that the scaling law proved for the one dimensional kinetic energy operator holds only approximately
in spaces with higher number of spatial dimensions.



\section*{Acknowledgements}

The Author would like to acknowledge many useful discussions with prof. J. Wosiek and M. Trzetrzelewski on the subject of this paper.


\small

\begin{thebibliography}{99}
\bibitem{dewit+hoppe+nicolai} B. de Wit, J. Hoppe, H. Nicolai, 'On the quantum mechanics of supermembranes', Nucl. Phys. B 305 (1988) 545
\bibitem{dewit+luscher+nicolai} B. de Wit, M. L\"uscher, H. Nicolai, 'The supermembrane is unstable', Nucl. Phys B 320 (1989) 135
\bibitem{banks+fischler+shenker+susskind} T. Banks, W. Fischler, S. Shenker, L. Susskind, 'M-theory as a matrix model: a conjecture', Phys. Rev. D 55 (1997) 6189
\bibitem{taylor} W. Taylor, 'M(atrix) theory: matrix quantum mechanics as a fundamental theory',Rev. Mod. Phys. 73 (2001) 419
\bibitem{claudson} M. Claudson, M.B. Halpern, 'Ground state wave functions', Nucl. Phys. B 250 (1985) 689-715
\bibitem{wosiek} J. Wosiek, 'Spectra of supersymmetric Yang-Mills quantum mechanics', Nucl. Phys. B 644 (2002) 85-112
\bibitem{korcyl3} P. Korcyl, 'Recursive approach to supersymmetric quantum mechanics for arbitrary fermion occupation number', Acta Phys. Pol. B 41 (2010) 795, arXiv: 0912.5265
\bibitem{wosiek1} J. Wosiek, 'Supersymmetric Yang-Mills quantum mechanics in various dimensions', Int. J. Mod. Phys. A 20 (2005) 4484-4491
\bibitem{wosiek2} J. Wosiek, 'On the $SO(9)$ structure of supersymmetric Yang-Mills quantum mechanics', Phys. Lett. B 619 (2005) 171-176
\bibitem{wosiek4} M. Campostrini, J. Wosiek, 'Exact Witten index in D=2 supersymmetric Yang-Mills quantum mechanics', Phys. Lett. B 550 (2002) 121-127
\bibitem{wosiek3} M. Campostrini, J. Wosiek, 'High precision study of the structure of D=4 supersymmetric Yang-Mills quantum mechanics', Nucl. Phys. B 703 (2004) 454-498
\bibitem{abramowitz} M. Abramowitz, I.A.Stegun, 'Handbook of Mathematical Functions with Formulas, Graphs, and Mathematical Tables', Dover Publications, New York, 1968
\bibitem{bateman} H. Bateman, 'Higher Transcendental functions, vol II', Bateman Manuscript Project, McGraw-Hill 1953
\bibitem{maciek2} M. Trzetrzelewski, J. Wosiek, 'Quantum systems in a cut Fock space',Acta Phys.Polon. B 35 (2004) 1615-1624, [hep-th/0308007]
\bibitem{maciek1} M. Trzetrzelewski, 'Quantum mechanics in a cut Fock space', Acta Phys. Polon. B 35 (2004) 2393-2416, [hep-th/0407059]
\bibitem{praca_magisterska} P. Korcyl, 'Classical trajectories and quantum supersymmetry', Phys. Rev. D 74 (2006) 115012
\bibitem{doktorat_macka} M. Trzetrzelewski, 'Supersymmetric Yang-Mills quantum mechanics with arbitrary number of colors', PhD thesis Jagiellonian University, 2006
\bibitem{hamermesh} M. Hamermesh, 'Group theory and its application to physical problems', Addison-Wesely, Readimng Mass, 1962
\bibitem{teoria_grup} J.-Q. Chen, J. Ping, F. Wang, 'Group representation theory for physicists', World Scientific, New Jersey, 2002
\bibitem{szego} G. Szeg\"o, 'Orthogonal polynomials', Vol. 23, 4th ed., Amer. Math. Soc. Colloq. Publ., Providence, RI, 1975
\bibitem{korcyl4} P. Korcyl, 'Exact solutions to $D=2$, Supersymmetric Yang-Mills Quantum Mechanics with $SU(3)$ gauge group', Acta Phys. Pol. B Proc. Suppl. 2 (2009) 623, arXiv: 0911.2152
\bibitem{korcyl1} P. Korcyl, 'Gauge invariant plane-wave solutions in supersymmetric Yang-Mills quantum mechanics', in preparation



%


%
%
%



\end{thebibliography}
\end{document}